\documentclass[12pt]{article}

\usepackage{amsmath}		
\usepackage{amssymb}

\usepackage{microtype}		
\usepackage{slashed}		
\usepackage{cite}			
\usepackage{hyperref}
\hypersetup{
	bookmarksnumbered,
	unicode,			
	colorlinks,			
	citecolor=[rgb]{.9,0,.5},	
	urlcolor=[rgb]{0,0,1},	
	linkcolor=[rgb]{0,.7,0}	
	}

\usepackage{empheq}
\usepackage[most]{tcolorbox}					
\definecolor{realcyan}{rgb}{0,1,1}
\definecolor{realyellow}{rgb}{1,1,0}
\newtcbox{\mymath}[1][]{%
    nobeforeafter, math upper, tcbox raise base,
    enhanced, colframe=realcyan!50!white,
    colback=realyellow!5!white, boxrule=1pt, drop lifted shadow, sharp corners
    #1}

\parskip=\medskipamount				

\numberwithin{equation}{section}

\mathcode`\*="702A                  	
\def\f#1#2{{\textstyle{#1\over#2}}}	

\catcode`@=11
\let\over\@@over
\catcode`@=12

\def\on#1#2{{\buildrel{{\mkern2.5mu\raise-.1em\hbox{$\scriptstyle#1$}\mkern-2.5mu}}\over{#2}}}
\def\ron#1#2{{\buildrel{{\raise-.1em\hbox{$\scriptstyle#1$}}}\over{#2}}}	
\def\ddt#1{\on{\hbox{\bf .\kern-1pt.}}#1}							

\providecommand{\smalltriangleright}{\vartriangleright}
\def\dd{\hbox{\large$\smalltriangleright$}}

\def\dig#1{\setbox0=\hbox{$#1M$}
	\hskip.06\wd0 \vrule width.07\wd0 height.63\wd0 depth.01\wd0
	\vrule width.37\wd0 height.63\wd0 depth-.56\wd0 \hskip-.4\wd0
	\vrule width.25\wd0 height.35\wd0 depth-.28\wd0
	\vrule width.07\wd0 height.35\wd0 depth-.17\wd0 \hskip.14\wd0}
\def\digamma{{\mathpalette\dig{}}}

\def\eq#1{\begin{equation}#1\end{equation}}

\def\rm{\mathrm}	

\def\={{\;=\;}}\def\+{{\;+\;}}
   \newcommand{\D}{\delta}

\newcommand{\bea}{\begin{eqnarray}}
\newcommand{\eea}{\end{eqnarray}}
\newcommand{\bref}[1]{(\ref{#1})}
\newcommand{\nn}{\nonumber}

\begin{document}
	
\hfill arXiv:1912.05092
\vskip-.1in
\hfill KEK-TH-2173
\vskip-.1in
\hfill YITP-SB-19-44

{\center
{\color{magenta}\fontsize{28pt}{34pt}\bf\sffamily
 T-dual Superstring Lagrangian with double zweibeins\\[.5in]

	 }

\href{mailto:mhatsuda@post.kek.jp}{Machiko Hatsuda} \\[.1in]
{\it
Department of Radiological Technology, Faculty of Health Science, Juntendo University\\
Hongo, Bunkyo-ku, Tokyo 113-0033, Japan\\
KEK Theory Center, High Energy Accelerator Research Organization\\
Tsukuba, Ibaraki 305-0801, Japan \\[.1in]
}
and \\[.1in]
\href{mailto:siegel@insti.physics.sunysb.edu}{Warren Siegel} \\[.1in]
{\it
\href{http://insti.physics.sunysb.edu/~siegel/plan.html}{C. N. Yang Institute for Theoretical Physics}\\
State University of New York, Stony Brook, NY 11794-3840}\\[.3in]

{\color[rgb]{0,.7,1}\today}\\[.5in]

}

{\abstract
We present superstring Lagrangians with manifest T-duality.
The Lagrangian version of the section conditions are necessary to make Lagrangians to be general  coordinate invariant.  
We show the general solution of section conditions. 
The D-dimensional left and right moving currents are the 2D-dimensional chiral current which causes the chiral boson problem.
We solve the problem by adding the unphysical 2D-dimensional anti-selfdual current with the selfduality constraints.
The Lagrange multipliers of the selfduality constraints play the role of the worldsheet zweibein
allowing the  Weyl invariant and Lorentz symmetric worldsheet.
Doubling the zweibein makes the type II $\kappa$-symmetry splitting into two sets of the type I $\kappa$-symmetries.}

\vfill

\thispagestyle{empty}
\newpage

{
\tableofcontents
}

\newpage
\section{Introduction}


T-duality is the characteristic feature of the string theory which leads to
a different picture from the Einstein gravity drastically at the short distance.
T-duality of string action is made manifest by doubling the spacetime coordinates 
\cite{Duff:1989tf,  Tseytlin:1990nb}.
Since physical string currents are the D-dimensional left and right currents,  
the physical current in the doubled spacetime is the
2D-dimensional chiral current which gives rise to the chiral boson problem \cite{Tseytlin:1990va}.
A Lagrangian only with chiral currents does not allow the conformal gauge which is useful for the quantum computation.
The Weyl invariance and the Lorentz covariance of the worldsheet are necessary especially for a superstring since the $\kappa$-symmetry involves the gauge transformation of the zweibein. 
Pasti, Sorokin and Tonin introduced a scalar
to resolve the chiral scalar problem  \cite{Pasti:1995tn,Pasti:1996vs}  which has a nonzero  vacuum  value leading to the spontaneous breaking down of the worldsheet symmetry.
It was applied to make manifest T-duality in \cite{Berman:2007xn}.

Bandos proposed the superstring Lagrangian with manifestly T-duality  
where the PST scalar field is used resulting double zweibeins \cite{Bandos:2015cha}.
The obtained superstring action has two sets of $\kappa$-symmetries
leading to simpler structure of the gauge invariance.
In our previous paper \cite{Hatsuda:2018tcx} 
doubling of the zweibein comes from the Lagrange multipliers
of the selfduality constraints for a bosonic string.
In this paper we extend it to the supersymmetric case.
Our Lagrangian is similar to the one obtained by Bandos \cite{Bandos:2015cha}.
Differences of our treatment in this paper from \cite{Bandos:2015cha} are the followings:
(1) We use both the selfdual and anti-selfdual currents which are extendable to non-abelian cases
instead of the Hodge dual of the selfdual current in \cite{Bandos:2015cha}.
(2) Lagrange multipliers of the selfduality constraints are used for double zweibeins,
instead of the PST scalar.
(3) The Wess-Zumino terms are written in bilinears of the currents as well as the kinetic term 
in our approach.
We also mention about our ``chiral" treatment in the previous paper \cite{Hatsuda:2015cia}.
The main differrence from \cite{Bandos:2015cha} is 
the dimensional reduction constraint
which is linear combination of the anti-selfdual currents.
Although we began by only the selfdual currents \cite{Hatsuda:2014qqa,Hatsuda:2014aza}, the dimensional reduction constraints 
involve the anti-selfdual currents leading to the Weyl invariant and Lorentz covariant worldsheet Lagrangian. 
As earlier studies many aspects of superstring Lagrangians with T-duality are examined such as
the NS/NS superstring
\cite{Blair:2013noa},
the doubled-yet-gauged spacetime formulation
\cite{Park:2016sbw,Borsato:2018idb,Sakamoto:2018krs}
and the pure spinor
\cite{Nikolic:2019wza}.

The string background is described by the gravity with the T-duality symmetry.
 It was shown that the classical gravity theory with manifest T-duality
is the same with the low energy 1-loop effective theory of the string    
\cite{Siegel:1993bj,Siegel:1993th,Siegel:1993xq} and 
the one for the chiral string in the $\alpha'$ order  \cite{Hohm:2013jaa}.
It is the gauge theory of gravity.
For the Einstein gravity the gauge generator and  the gauge field are  the momentum $p_{\rm M}$ and  the vielbein
$e_{\rm A}{}^{\rm M}$ which make the covariant derivative $\nabla_{\rm A}= e_{\rm A}{}^{\rm M}p_{\rm M}$.
For the stringy gravity the momentum includes the winding mode 
$\dd_M(\sigma)=(p_{\rm M},\partial_\sigma x^{\rm M})$ with the D-dimensional index ${\rm M}$ and 
the 2D-dimensional index $M$.
They satisfy the affine Lie algebra
whose consistency requires the nondegenerate group metric $\eta_{MN}$.
This affine Lie algebra is realized by the 2D-dimensional canonical coordinates $P_M$ and $X^M$. 
The covariant derivative is extended to $\nabla_{\rm A} \to \dd_{A}(\sigma)=E_A{}^M\dd_M$
with the vielbein $E_A{}^M(X)$. 
The gauge transformation of the vielbein 
under the general coordinate transformation 
is given by  $\delta E_A{}^M=dE_A{}^M+{\cal L}_\Lambda E_A{}^M$
with the differential term $dE_A{}^M$ and the ``new" Lie derivative ${\mathcal L}_\Lambda E_A{}^M$.
The  ``new" Lie derivative gives rise to the O(D,D) transformation
\eq{\delta E_A{}^M=E_A{}^N(-\partial_N\Lambda^M+\partial^M\Lambda_N)
	~~,~~
\delta E_M{}^A=	(\partial_M\Lambda^N-\partial^N\Lambda_M)E_N{}^A~~~. \label{EMA}}
Throughout  this paper the 2D-dimensional index $M$ is raised and lowered by the O(D,D) invariant metric $\eta^{MN}$.
The tangent vector is transformed as O(D,D) 
\eq{		\displaystyle\frac{\partial}{\partial X'{}^M}=~\displaystyle\frac{\partial X{}^N}{\partial X'{}^M}\frac{\partial}{\partial X^N} =~(\delta_M^N+
	\partial_M\Lambda^N)\partial_N\approx~(\delta_M^N+\partial_M\Lambda^N-\partial^N\Lambda_M)\partial_N	~~~}
where  the section conditions
on arbitrary functions $\Psi_I(X)$ with $I=1,2,\cdots$ are used
\eq{\eta^{MN}\partial_M \partial_N \Psi_I(X)=\eta^{MN}\partial_M \Psi_I(X)\partial_N \Psi_J(X)=0\label{sectionHam}~~~.}
The cotangent vector should be also transformed as O(D,D)
\eq{dX'{}^M=~dX^N\displaystyle\frac{\partial X'{}^M}{\partial X^N} =~dX^N(\delta_N^M-
	\partial_N\Lambda^M)
	\approx~dX^N(\delta_N^M-\partial_N\Lambda^M+\partial^M\Lambda_N)\label{GCT}~~~,}	
so the Lagrangian version of the section conditions are necessary as shown in our previous paper \cite{{Hatsuda:2018tcx}}
\eq{dX^M\eta_{MN}dX^N=dX'{}^M\eta_{MN}
	dX'{}^N=0=dX^M \eta_{MN} \partial_L\Lambda^N~~~.\label{Lagsec}} 
These conditions guarantee the consistency for the cotangent vector $dX^M$ and the tangent vector
$\frac{\partial}{\partial X^N}$ as $\langle dX^M , \frac{\partial}{\partial X^N}\rangle =\delta_N^M$. 
The Lagrangian version of the section conditions also guarantee the 
coordinate invariance of the currents in  curved backgrounds
\eq{
	J_m{}^A=\partial_m X^ME_M{}^A~,~
	\delta J_m{}^A=
	(\partial_m \delta X^M)E_M{}^A+\partial_m X^M \delta E_M{}^A=0 ~~~,}
where index ${m}$ runs $\tau,\sigma$ for a string and $\tau$ for a particle.
Then the Lagrangian is made to be coordinate invariant.

The organization of the paper is the following:
In the next section we present solutions of the section conditions 
in Hamiltonian \bref{sectionHam} and in Lagrangian \bref{Lagsec} explicitly.
In section 3, the worldsheet gauges are examined where the Lagrange multipliers 
of the Virasoro constraints and the selfduality constraints in Hamiltonian 
become double zweibeins.
In section 4  the Virasoro constraints and the selfduality constraints in a non-abelian space are obtained. 
We begin by the Hamiltonian formalism where the covariant derivative commutes with the symmetry generator.
The covariant derivative is the selfdual current while the orthogonal transformed 
symmetry generator becomes the anti-selfdual current.
The $\sigma$-diffeomorphism Virasoro operator includes the anti-selfdual current,
so the $\sigma$ derivative computed by the canonical commutator coincides with the one computed by the chain rule differential.
Including the anti-selfdual current is a similar formulation given 
in \cite{Sen:2019qit}
where the Lagrangian for selfdual 2n-form fields
is written with the anti-seldfual form.

In section 5, the superstring Lagrangians with manifest T-duality are presented.  
For two sets of nondegenerate superalgebras the selfdual and anti-selfdual currents are given concretely.
The Hamiltonian of the superstring includes the Virasoro constraints, the selfduality constraints and the dimensional reduction constraints for unphysical fermions.
The Lagrangian with double zweibeins makes the type II $\kappa$-symmetry to be two sets of the type I $\kappa$-symmetries leading to simpler computation.   
We also show 
how to reduce to the Green-Schwarz superstring action by gauge fixing and sectioning.


\section{Section conditions}

\subsection{Section conditions in Hamiltonian}

The manifestly T-duality space is defined by the string current algebra,
where the worldsheet spatial diffeomorphism is suppressed consistently as the section conditions.
The Virasoro operators are 
\eq{
	{\renewcommand{\arraystretch}{1.6}
		\left\{\begin{array}{ccl}
			{\cal H}_\tau&=&\f12P_M G^{MN}P_N\\
			{\cal H}_\sigma&=&\f12P_M{\eta}^{MN}P_N
		\end{array}\right.}\label{VirasoroP}}
where $G^{MN}$ is the O(D,D) gravitational background metric while ${\eta}^{MN}$ is the O(D,D) invariant  metric. 
${\cal H}_\tau$ is the Hamiltonian in the conformal gauge.
${\cal H}_\sigma=0$ is realized on arbitrary  fields  $\Psi_i(X^M)~,~_{i=1,2,\cdots}$ as weak and strong section conditions
\eq{
	\eta^{MN}\partial_M\partial_N\Psi_i(X)=\eta^{MN}\partial_M\Psi_i(X)\partial_N\Psi_j(X)=0 ~~
	\rm{for}~{i,j=1,2,\cdots .}
}
The Fourier transformation introduces momenta $P_{i;M}$ for each function as
\eq{\Psi_i(X^M)=\int d^{2D}P_i e^{-iP_{i;M}X^M}\tilde{\Psi}_i(P_{i;M})~~~.\label{Fourier}} 
The weak section condition gives 
\eq{\eta^{MN}\partial_M\partial_N\Psi_i(X)=0~
	\to~
	\eta^{MN}	P_{i}{}_{M}P_{i;N}=0~~\rm{for}~{i=1,2,\cdots .}\nn} 
Let us divide the 2D directions of $P_M$ into two Euclidean D-dimensional directions of
the positive metric $\bar{P}_{\overline{M}}$ and the one of the negative metric $\underline{P}_{\underline{M}}$
\eq{\eta^{MN}P_{i;M}P_{i;N}=\sum_{\overline{M}=1}^D (\bar{P}_{i;\overline{M}})^2-\sum_{\underline{M}=1}^D (\underline{P}_{i;\underline{M}})^2=0
	\to \sum_{\overline{M}=1}^D (\bar{P}_{i;\overline{M}})^2=	\sum_{\underline{M}=1}^D (\underline{P}_{i;\underline{M}})^2\equiv	|P_{i}|^2}
The strong section condition gives
\eq{\eta^{MN}\partial_M\Psi_i(X)\partial_N\Psi_j(X)=0~\to  ~\eta^{MN}P_{i;M}P_{j;N}=0~~\rm{for}~{i\neq j=1,2,\cdots}}
which leads to
\eq{\eta^{MN}P_{i;M}P_{j;N}=\displaystyle{\sum_{\overline{M}=1}^D} \bar{P}_{i;\overline{M}} \bar{P}_{j;\overline{M}}-\sum_{\underline{M}=1}^D \underline{P}_{i;\underline{M}} \underline{P}_{j;\underline{M}}=	|P_{i}||P_{j}|(\rm{cos} \bar{\theta}_{ij}-\rm{cos} \underline{\theta}_{ij})=
	0~\rm\nn}
\eq{	\Rightarrow
	\bar{ \theta}_{ij}= \underline{\theta}_{ij}~~~.}
In order to coincide all angles of infinite number of arbitrary vectors, $	\bar{ \theta}_{ij}= \underline{\theta}_{ij}
$, the positive and negative vectors  $\bar{P}_{i;\overline{M}}$ and  $\underline{P}_{i;\underline{M}}$ must be equal
up to an O(D) rotation $A_{\overline{M}}{}^{\underline{N}}$ 
\eq{ \bar{P}_{i;\overline{M}}=A_{\overline{M}}{}^{\underline{N}}\underline{P}_{i;\underline{N}}~~{\rm{for}}~i=1,2,\cdots~,~
	A_{\overline{M}}{}^{\underline{N}}A_{\overline{M}}{}^{\underline{L}}=\delta^{\underline{NL}}~,~
	A_{\overline{M}}{}^{\underline{N}}A_{\overline{L}}{}^{\underline{N}}=\delta_{\overline{M}\overline{L}}~~~.
	\label{solution}}
The infinite number of vectors $\bar{P}_{i;\overline{M}}$ and $\underline{P}_{i;\underline{M}}$ 
are recognized as infinite number of points in the momentum coordinate space  
$\bar{P}_{\overline{M}}$ and $\underline{P}_{\underline{M}}$.
Therefore the reducibility \bref{solution} eliminates a half space as
\eq{ \bar{P}_{\overline{M}}=A_{\overline{M}}{}^{\underline{N}}\underline{P}_{\underline{N}}~~\to~~
	0=
	{\renewcommand{\arraystretch}{1.6}
		\left(	\begin{array}{c|c}~{\bf 1}_{\overline{M}}{}^{\underline{N}}~&
			-A_{\overline{M}}{}^{\underline{N}}\\\hline 0&0	\end{array}	\right)
		\left(\begin{array}{c}\bar{P}_{\overline{N}}\\\hline \underline{P}_{\underline{N}}\end{array}\right)}~~~.\label{reducible}}

Interchanging  $\bar{P}_{\overline{D}}$ and $\underline{P}_{\underline{D}}$ of
 the positive and negative vectors
 and renaming them as ${P}_{\underline{0}}$ and ${P}_{\overline{0}}$ 
 make  the Lorentz covariant left and right vectors.
The  interchanging matrix ${\cal I}$ makes  the left/right vectors $P_{\rm{ L/R}}=({P}_{\overline{M}},{P}_{\underline{M}})$ from
the positive/negative vectors   $P_{\rm{P/N}}=(\bar{P}_{\overline{M}},\underline{P}_{\underline{M}})$.
It also makes the O(D,D) invariant metric to be diagonal $(\eta_{\overline{MN}}, -\eta_{\overline{MN}})$
from diagonal $({\bf 1},-{\bf 1})$ 
\bea
&{\cal I}=	{\renewcommand{\arraystretch}{0.9}
\left(\begin{array}{cc|cc}{\bf 1}&{\bf 0}&{\bf 0}&{\bf 0}\\{\bf 0}& 0&{\bf 0}&1\\\hline
	{\bf 0}&{\bf 0}&{\bf 1}&{\bf 0}\\{\bf 0}&1&{\bf 0}&	0\end{array}	\right)}
~,~P_{\rm{L/R}}={\cal I}P_{\rm{P/N}}~,~
\eta_{MN}=
{\cal I}	{\renewcommand{\arraystretch}{1.3}
		\left(\begin{array}{c|c}{\bf 1}&{\bf 0}\\\hline{\bf 0}&-{\bf 1}\end{array}	\right)
		{\cal I}
		=
		\left(\begin{array}{c|c}{\eta}_{\overline{M}\overline{N}}&0\\\hline0&
			-{\eta}_{\overline{M}\overline{N}}\end{array}	\right)}&\nn\\
	\label{Minkowski} 
\eea
Plugging \bref{Minkowski}  into the reducibility relation
\bref{reducible} the left/right momenta also satisfy the reducibility condition with 
a new reducibility matrix denoted by the same matrix $A$  as  
\eq{	P_{\overline{M}}=A_{\overline{M}}{}^{\underline{N}}P_{\underline{N}}\label{PMP}}
The matrix $A_{\overline{M}}{}^{\underline{N}}$ is an element of 
O(D$-$1,1) group with respect to  ${\eta}_{\overline{M}\overline{N}}$ and ${\eta}^{\overline{M}\overline{N}}$ as 
$A{\eta} A^T={\eta}$ and $A^T{\eta} A={\eta}$.
By using $A_{\overline{M}}{}^{\underline{N}}$ we fix the left and  right Lorentz symmetries, 
${S}_{\overline{M}\overline{N}}$ and ${S}_{\underline{MN}}$, as
\eq{
	{S}_{\rm{physical};\rm{MN}}={S}_{\overline{M}\overline{N}}-A_{\overline{M}}{}^{\underline{L}}
	A_{\overline{N}}{}^{\underline{K}} 	{S}_{\underline{LK}}~~,~~
	0={S}_{\overline{M}\overline{N}}+A_{\overline{M}}{}^{\underline{L}}A_{\overline{N}}{}^{\underline{K}}
	{S}_{\underline{LK}}
	\label{fixLorentz} 	}
It is denoted that the left and the right Lorentz generators satisfy the same Lorentz algebras with the opposite signatures as
$[S_{\rm{left}},S_{\rm{left}} ]=S_{\rm{left}}$ and 
$[S_{\rm{right}},S_{\rm{right}} ]=-S_{\rm{right}}$. 
Linear combination of the left/right momenta
brings to the conventional basis $p_{\rm{M}}$ and $p^{\rm{M}}$ as
\eq{{\renewcommand{\arraystretch}{1.6}	\left\{
		\begin{array}{ccl}
			p_{\rm{physical};\rm{M}}&=&\f12(P_{\overline{M}}+A_{\overline{M}}{}^{\underline{L}}P_{\underline{L}})=P_{\overline{M}}\\
			p_{}{}^{\rm{M}}&=&\f12(P_{\overline{M}}-A_{\overline{M}}{}^{\underline{L}}P_{\underline{L}})=0
		\end{array}\right.	}}
The Fourier functions have the form of $\tilde{\Psi}_i(P_{i;M})=\tilde{\Psi}_i(p_{i;\rm{M}},p_{i}{}^{\rm{M}}=0)$.
The Fourier integration with respect to $p_{i}{}^{\rm{M}}$
gives $\int d^Dp_{i}{}^{\rm{M}}e^{-ip_i{}^{\rm{M}}y_{\rm{M}} }\tilde{\Psi}_i(P_{M})=\delta({y_{\rm{M}}})\tilde{\Psi}_i(p_{\rm M})$, 
so the resultant functions are 
\eq{
	{\Psi}_i(X^M)=\Psi_i(x^{\rm{M}},y_{\rm{M}}=0)\label{316}~~~.
}
The half of 2D coordinates are suppressed. 
Other solutions such as $x^{\rm{M}}=0,~y_{\rm M}\neq 0$, are obtained by regular O(D,D) matrix transformations.

\subsection{Section conditions in Lagrangian}

The Lagrangian version of the section conditions are given  in our previous paper\cite{Hatsuda:2018tcx}.
We also showed that the Virasoro operators \bref{VirasoroP} become
the line element and the constraint in Lagrangian formalism 
\eq{{\renewcommand{\arraystretch}{1.6}
		\left\{\begin{array}{cl}
			\f12d{X}^MG_{MN}d{X}^N&=
			d^2s\\
			\f12d{X}^M{\eta}_{MN}d{X}^N
			&=0
		\end{array}\right.}\label{Lagversec}} 
The second line is the section condition suppressing the degrees of freedom generated by the  $\sigma$-diffeomorphism which corresponds to the weak section condition. 
Under an infinitesimal coordinate transformation $X^M\to X'{}^M=X^M-\Lambda^M(X)$
the additional condition is required \cite{Hatsuda:2018tcx} 
corresponding to the strong section condition
\eq{	d{X}^M{\eta}_{MN}\partial_L{\Lambda}^N=0\label{Lagversecstr}}

The Lagrangian version of the section conditions are solved analogously  in \bref{PMP}
\eq{\eta_{MN}d{X}^Md{X}^N=\eta_{MN}d{X}'{}^Md{X}'{}^N=0~,~\eta_{MN}d{X}^Md{X'}^N=0~\to
	~d{X}^{\overline{M}}	=d{X}^{\underline{N}}	(A^{-1}){}_{\underline{N}}{}^{\overline{M}}  \label{dXdXA}}
The left and right Lorentz symmetries are fixed analogously to \bref{fixLorentz}.
The conventional coordinates are introduced as
\eq{{\renewcommand{\arraystretch}{1.6}\left\{
		\begin{array}{ccl}
			{x}^{\rm M}&=&{X}^{\overline{M}}+{X}^{\underline{N}}(A^{-1}){}_{\underline{N}}{}^{\overline{M}}\\
			{y}_{\rm M}&=&{X}^{\overline{M}}-{X}^{\underline{N}}(A^{-1}){}_{\underline{N}}{}^{\overline{M}}
		\end{array}
		\right.}\label{322}}
A solution of the Lagrangian version of the weak section condition \bref{dXdXA} is given by
\eq{{\renewcommand{\arraystretch}{1.6}\left\{
		\begin{array}{ccl}	
			dx^{\rm M}&=&d{X}^{\overline{M}}+d{X}^{\underline{N}}(A^{-1}){}_{\underline{N}}{}^{\overline{M}} =2dX^{\overline{M}}\\
			dy_{\rm M}&=&d{X}^{\overline{M}}
			-d{X}^{\underline{N}}
			(A^{-1}){}_{\underline{N}}{}^{\overline{M}} =0
		\end{array}\right.}
	~\to~d{X}^{M}=(d{x}^{\rm{M}},d{y}_{\rm{M}}=0)~~~.}
This solution supplements the solution of $y_{\rm M}=0$ in \bref{316}.
The weak condition in \bref{Lagversecstr} is examined as 
$dX^M\eta_{MN}\partial_L\Lambda^N=0=dx^{\rm M}\frac{\partial}{\partial x^L}\Lambda_{\rm M}=0
$ leading to $\Lambda_{\rm M}=$constant,
where $\frac{\partial}{\partial y_{\rm M}}\Lambda=0$ and $dy_{\rm M}=0$ are used. 
So the transformed coordinate  is still solution
$dy'_{\rm M}=dy_{\rm M}-d\Lambda_{\rm M}=0$.
Other solutions are obtained by regular O(D,D) matrix transfomations.

\section{Worldsheet gauges}

The Lagrange multipliers of the Virasoro constraints in  Hamiltonian 
are zweiben gauge fields in Lagrangian.
The choice of the zweibein gauge links the target space symmetry.
In order to focus on this relation 
we consider a bosonic string in a D-dimensional flat space without the $B$-field.
The D-dimensional target space coordinate and the canonical conjugate are denoted by
$(x^{\rm M},p_{\rm M})$ with the usual Lorentz metric $\eta_{\rm{MN}}$. 
The Hamiltonian for a string in the D-dimensional flat space gives 
the Lagrangian in the Weyl-Lorentz gauge of the zweibein
\bea
H_0&=&g_-\f14(p+\partial_\sigma x)_{\rm M}{}^2+g_+\f14(p-\partial_\sigma x)_{\rm M}{}^2
\nn\\
L_0&=&\partial_\tau x^{\rm M}~p_{\rm M}-H_0~=~
\frac{1}{e}~(e_+{}^m\partial_m{x}^{\rm M})~(e_-{}^n\partial_n x^{\rm N})~\eta_{\rm{MN}}
\nn\\
&&{\renewcommand{\arraystretch}{1.3}\left\{	\begin{array}{ccl}
		e_a{}^m&=&\left(	\begin{array}{cc}e_-{}^\tau&e_-{}^\sigma\\e_+{}^\tau& e_+{}^\sigma\end{array}	\right)=	\left(		\begin{array}{cc}1&-g_-\\1& g_+\end{array}	\right)
		\label{epmzwe}\\
		e&=&\rm{det}~e_a{}^m={g_++g_-}\end{array}
	\right.}\label{pmHL}
\eea

In the manifestly T-duality formulation the D-dimensional left/right coordinates are 
treated as independent 2D-dimensional coordinates;
the coordinates  $X^M=(X^{\overline{M}},~X^{\underline{M}})$ 
and the conjugate momenta
$P_M=(P_{\overline{M}},P_{\underline{M}})$ .
On the other hand the 2D-dimensional left/right moving currents, $P_M\pm \partial_\sigma X^N\eta_{NM}$,~
include not only the D-dimensional left/right moving currents  $P_M+\partial_\sigma X{}^N\eta_{NM}=\left(({P}+\partial_\sigma{X})_{\overline{M}},~({P}-\partial_\sigma{X})_{\underline{M}}\right)
$  but also unphysical currents
$P_M-\partial_\sigma X{}^N\eta_{NM}=\left(({P}-\partial_\sigma{X})_{\overline{M}},~({P}+\partial_\sigma{X})_{\underline{M}}\right)
$. 
$P_M+\partial_\sigma X{}^N\eta_{NM}$ is the selfdual current and $P_M-\partial_\sigma X{}^N\eta_{NM}$ is the anti-selfdual currents.

We review our chiral approach \cite{Hatsuda:2015cia}. The Hamiltonian includes only physical currents.
The selfduality constraint is imposed by the linear combination of the left/right anti-selfdual currents
in such a way that the stringy anomaly is cancelled, $(P-\partial_\sigma X)_{\overline{M}}-
(P+\partial_\sigma X)_{\underline{M}}=0$.
The chiral Hamiltonian with the linear selfduality constraint is given by
\bea
H_{\rm{chiral}}&=&g_-\f14(P+\partial_\sigma X)_{\overline{M}}{}^2+g_+\f14(P-\partial_\sigma X)_{\underline{M}}{}^2\nn\\
&&+\mu^{\rm{ M}}\left\{(P-\partial_\sigma X)_{\overline{M}}-
(P+\partial_\sigma X)_{\underline{M}}\right\}
\nn\\
L_{\rm{chiral}}&=&\partial_\tau X^{M}~P_{M}-H_{\rm{chiral}}~\nn\\
&=&\frac{1}{g_-}~\left\{(e_-{}^m\partial_m{X}^{\overline{M}}-\mu^{\rm M})^2 +g_-\partial_\sigma X^{\overline{M}}
(e_-{}^m\partial_m{X}^{\overline{M}})\right\}\nn\\
&&+\frac{1}{g_+}~\left\{(e_+{}^m\partial_m{X}^{\underline{M}}+\mu^{\rm M})^2 -g_+\partial_\sigma X^{\underline{M}}(e_+{}^m\partial_m{X}^{\underline{M}})\right\}
\eea
where the zweibein $e_a{}^m$ is given in \bref{pmHL}.
After integrating out the Lagrangian multiplier $\mu^{\rm{M}}$ and 
rewriting in terms of the usual coordinates $x^{\rm{M}}$ and $y_{\rm{M}}$ in \bref{322}, 
the worldsheet covariant Lagrangian is obtained
\bea
L_{\rm{chiral}}&=&\frac{1}{e}~(e_+{}^m\partial_m{x}^{\rm M})~(e_-{}^n\partial_n x^{\rm N})~\eta_{\rm MN}
-\epsilon^{mn}\partial_m x^{\rm M}\partial_n y_{\rm M}~~~.
\eea
The first term is both  D-dimensional and worldsheet covariant kinetic term. 
The second term is total derivative in the bosonic case, but it contributes to the supersymmetric case.

Next let us include the anti-selfdual currents.
The Hamiltonian for a string in the 2D-dimensional flat space gives 
the Lagrangian in the Weyl-Lorentz gauge with two zweibeins:
\bea
H_1&=&g_-\f14({P}+\partial_\sigma{X})_{\overline{M}}{}^2+g_+\f14({P}-\partial_\sigma{X})_{\underline{M}}{}^2\nn\\
&&+(g_++\lambda_+)\f14({P}-\partial_\sigma{X})_{\overline{M}}{}^2+(g_-+\lambda_-)\f14({P}+\partial_\sigma{X})_{\underline{M}}{}^2\nn\\
L_1&=&\partial_\tau X^M~{P}_M-H_1\nn\\
&=&	\frac{1}{\bar{e}}~(\bar{e}_+{}^m\partial_m{X}^{\overline{M}})(\bar{e}_-{}^n\partial_n{X}^{\overline{N}})\eta_{\overline{M}\overline{N}}+
\frac{1}{\underline{e}}~(\underline{e}_+{}^m\partial_m{X}^{\underline{M}})(\underline{e}_-{}^n\partial_n{X}^{\underline{N}})\eta_{\overline{M}\overline{N}}
\label{ebareul}\\
&&{\renewcommand{\arraystretch}{1.3}\left\{	\begin{array}{ccl}
		\bar{e}_{a}{}^m
		&	=&	\left(		\begin{array}{cc}1&-g_-\\1&g_++\lambda_+\end{array}	\right)\\
		\bar{e}&=&
		e+\lambda_+\end{array}
	\right.}~,~
{\renewcommand{\arraystretch}{1.3}\left\{	\begin{array}{ccl}
		\underline{e}_{a}{}^m
		&	=&	\left(		\begin{array}{cc}1&-(g_-+\lambda_-)\\1&g_+\end{array}	\right)\\
		\underline{e}&=&
		e+\lambda_-\end{array}
	\right.}\nn
\eea
Two zweibeins and two worldsheet coordinates allow two independent worldsheets,
so  $L_1$ is sum of the left and the right sectors.
It will be convenient to calculate the $\kappa$-symmetry invariance as shown in the next section.

The Lagrangian $L_1$ is rewritten in such a way that 
the kinetic term becomes 2D-dimensional covariant,
\bea
L_2&=&\frac{1}{{e}}~({e}_+{}^m\partial_m{X}^{{M}})({e}_-{}^n\partial_n{X}^{{N}})\hat{\eta}_{{M}{N}}
\nn\\
&&+	(\frac{1}{\bar{e}}-\frac{1}{{e}} )~({e}_-{}^m\partial_m{X}^{\overline{M}})^2+
(\frac{1}{\underline{e}}-\frac{1}{e})~({e}_+{}^m\partial_m{X}^{\underline{M}})^2 \label{L2}
\eea
where the zweibein $e_a{}^m$ is the same one in \bref{pmHL}.
In the conformal gauge $g_{\pm}=1$ $L_2$ becomes
\eq{L_2=\f12\left\{(\partial_+ X^M)(\partial_- X^N)\hat{\eta}_{MN}
	-\frac{\lambda_+}{2+\lambda_+}(\partial_-X^{\overline{M}})^2
	-\frac{\lambda_-}{2+\lambda_-}(\partial_+X^{\underline{M}})^2\right\}
}
with $\partial_\pm=\partial_\tau\pm\partial_\sigma$.
The selfdual constraints obtained by varying $\lambda_\pm$  in terms of $x^{\rm M},~y_{\rm M}$
in \bref{322}
are squares of the usual selfduality condition $\partial_m y_{\rm M}=\epsilon_{mn}\partial^n x^{\rm M}$  equivalently $\partial_m (X\eta)^M=\epsilon_{mn}\partial^n X^{M}$ as
\eq{	\left\{	{\renewcommand{\arraystretch}{1.3}	\begin{array}{ccl}
			\left(	\partial_-X^{\overline{M}}\right)^2&=&\f14\left((\partial_\tau x^{\rm M}-\partial_\sigma y_{\rm M})
			+(\partial_\tau y_{\rm M}		-\partial_\sigma x^{\rm M})	\right)^2=0	\\
			\left(	\partial_+X^{\underline{M}}\right)^2&=&\f14\left((\partial_\tau x^{\rm M}-\partial_\sigma y_{\rm M})		-(\partial_\tau y_{\rm M}		-\partial_\sigma x^{\rm M})\right)^2=0
	\end{array}}\right.}
The D-dimensional dual coordinate $dy_{\rm M}$ is solved in terms of $dx^{\rm M}$,
if the selfduality constraint is solved.
But squares of the selfduality constraints are weaker, and it contributes to make the worldsheet covariance manifest.
The Lagrangian $L_2$ is rewritten in terms of $x^{\rm M},~y_{\rm  M}$ in \bref{322} as
\bea
L_2&=&
\frac{1}{{e}}\left\{({e}_+{}^m\partial_m{x}^{\rm{M}})({e}_-{}^n\partial_n{x}^{\rm{N}}){\eta}_{\rm{MN}}
+({e}_+{}^m\partial_m{y}_{\rm{M}})({e}_-{}^n\partial_n{y}_{\rm{N}}){\eta}^{\rm{MN}}\right.\nn\\
&&~~~\left.+\frac{\lambda_+}{e+\lambda_+}(\partial_\tau x^{\rm M}-\partial_\sigma y_{\rm M})
(\partial_\sigma x^{\rm M}-\partial_\tau y_{\rm M})\right\}
\eea
in the gauge $\frac{1}{\bar{e}}+\frac{1}{\underline{e}}=\frac{2}{e}$ which is 
$\lambda_-=\frac{e\lambda_+}{e+2\lambda_+}$.

The Lagrangian $L_1$ is further rewritten in such a way that 
both the kinetic term and the constraints become 2D-dimensional covariant:
\bea
L_3&=&\frac{1}{\hat{e}}~(\hat{e}_+{}^m\partial_m{X}^{{M}})(\hat{e}_-{}^n\partial_n{X}^{{N}})\hat{\eta}_{{M}{N}}
+	\frac{1}{\hat{\lambda}}~(\hat{\lambda}_+{}^m\partial_m{X}^{{M}})
(\hat{\lambda}_-{}^n\partial_n{X}^{{N}}){\eta}_{{M}{N}}\\
&&{\renewcommand{\arraystretch}{1.6}\left\{	\begin{array}{ccl}
		\hat{e}_{a}{}^m
		&	=&	\left(	{\renewcommand{\arraystretch}{1.3}	\begin{array}{cc}1&-\hat{g}_-\\1&\hat{g}_+\end{array}	}\right)\\
		\hat{g}_{+}&=&g_++
		\displaystyle\frac{\lambda_+\underline{e}}{\bar{e}+\underline{e}}\\
		\hat{g}_{-}&=&g_-+\displaystyle\frac{\lambda_-\bar{e}}{\bar{e}+\underline{e}}\\
		\hat{{e}}&=&\displaystyle 2(\frac{1}{\bar{e}}+\frac{1}{\underline{e}})^{-1}
	\end{array}\right.
	~,~		\left\{	\begin{array}{ccl}
		\hat{\lambda}_{a}{}^m
		&	=&	\left(		{\renewcommand{\arraystretch}{1.3}	\begin{array}{cc}1&-\hat{\lambda}_-\\1&\hat{\lambda}_+\end{array}	}\right)\\
		\hat{{\lambda}}_{+}&=&g_++
		\displaystyle\frac{(2\bar{e}-e)\underline{e}}{-\bar{e}+\underline{e}}\\
		\hat{{\lambda}}_{-}&=&g_-+
		\displaystyle\frac{e\bar{e}}{-\bar{e}+\underline{e}}\\
		\hat{{\lambda}}&=&\displaystyle 2(\frac{1}{\bar{e}}-\frac{1}{\underline{e}})^{-1}
	\end{array}\right.}\nn
\eea
The conformal gauge in the first term is given by
\eq{ 	\left\{		{\renewcommand{\arraystretch}{1.3}	\begin{array}{ccl}
			\lambda_+&=&e\displaystyle\left(\frac{1}{g_+}-1 \right)\\
			\lambda_-&=&e\displaystyle\left(\frac{1}{g_-}-1 \right)\end{array}	}\right.
	~\to~\hat{e}_a{}^m=
	\left(		{\renewcommand{\arraystretch}{1.3}	\begin{array}{cc}1&-1\\1&1\end{array}	}\right)
	~~~.
}
In this gauge the second term does not allow the conformal gauge, $\hat{\lambda}_\pm=1$, 
since they are originally selfduality constraints,
\eq{ 	\left\{		{\renewcommand{\arraystretch}{1.6}	\begin{array}{ccl}
			\hat{\lambda}_+&=&\displaystyle\frac{2(g_++g_--g_+g_-)}{g_+-g_-}\\
			\hat{\lambda}_-&=&\displaystyle\frac{2g_+g_-}{g_+-g_-}\end{array}	}\right.
	~\leftrightarrow~
	\left\{		{\renewcommand{\arraystretch}{1.3}	\begin{array}{ccl}
			g_+&=&\displaystyle\frac{2\hat{\lambda}_-}{\hat{\lambda}_++\hat{\lambda}_-   -2}\\
			g_-&=&\displaystyle\frac{2\hat{\lambda}_-}{\hat{\lambda}_++\hat{\lambda}_-   +2}\end{array}	}\right.	~~~.}
This is in contrast to $L_2$ in \bref{L2} where the conformal gauge in the zweibein $e_a{}^m$ is allowed 
for both the kinetic term and the selfduality constraints.
The solutions of the selfduality constraints and the section condition
lead to the same physical degrees of freedom. 
The second term should be imposed as the section condition which is 
the 2D-dimensional covariant orthogonal condition,
\eq{(\hat{\lambda}_+{}^m\partial_m{X}^{{M}})
	(\hat{\lambda}_-{}^n\partial_n{X}^{{N}}){\eta}_{{M}{N}}=0~~~.}
A solution of the section condition is given in \bref{322}.

\section{Non-abelian space currents}

\subsection{Algebra and currents}
A nondegenerate graded Lie algebra generated by  $G_{I}$ 
has the nondegenerate group metric $\eta_{IJ}$ and the totally graded antisymmetric 
structure constant $f_{IJK}$
\eq{[G_I,G_J\}=if_{IJ}{}^KG_K~,~\rm{tr}(G_IG_J)=\eta_{IJ}=\f12\eta_{(IJ]}
	~,~f_{IJK}\equiv f_{IJ}{}^L\eta_{LK}=\frac{1}{3!}f_{[IJK)} \label{GGfG}}
with the graded bracket 
$[A,B\}=AB-(-)^{AB}BA$
and the graded symmetrized and antisymmetrized indices   
$(A,B]=AB+(-)^{AB}BA$ and $[A,B)=AB-(-)^{AB}BA$.
A group element $g(Z)$ with coordinates $Z^I$ gives two kinds of currents and derivatives:
The left-invariant current $J^I$, the particle covariant derivative $\nabla_I$, the right-invariant current $\tilde{J}^I$ and the particle symmetry generator $\tilde{\nabla}_I$ are given by
\eq{{\renewcommand{\arraystretch}{1.6}
		\begin{array}{lccl}
			\rm{Left}{\mathchar`-}\rm{invariants}:&	g^{-1}dg=iJ^IG_I=idZ^MR_M{}^IG_I&,&\nabla_I=(R^{-1})_I{}^M\frac{1}{i}\displaystyle\frac{\partial}{\partial Z^M}\\
			\rm{Right}{\mathchar`-}\rm{invariants}:&dgg^{-1}=i\tilde{J}^IG_I=idZ^ML_M{}^IG_I&,&\widetilde{\nabla}_I=(L^{-1})_I{}^M\displaystyle\frac{1}{i}\frac{\partial}{\partial Z^M}
\end{array}}  }	
They satisfy the following Maurer-Cartan equations and the algebras
\eq{{\renewcommand{\arraystretch}{1.6}
		\begin{array}{llcl}
			\rm{Left}{\mathchar`-}\rm{invariants}:&dJ^I=\f12J^J\wedge J^Kf_{KJ}{}^I&,&
			[{\nabla}_I,{\nabla}_J\}=-if_{IJ}{}^K{\nabla}_K
			\\
			\rm{Right}{\mathchar`-}\rm{invariants}:&d\tilde{J}^I=-\f12\tilde{J}^J\wedge \tilde{J}^K f_{KJ}{}^I	&,&
			[\widetilde{\nabla}_I,\widetilde{\nabla}_J\}=if_{IJ}{}^K\widetilde{\nabla}_K\\
			\rm{Mixed}&&&[{\nabla}_I,\widetilde{\nabla}_J\}=0 			\end{array}}\label{nab}}
The left- and the right-invariant currents are related by the orthogonal matrix $M_I{}^J$ as
\cite{Hatsuda:2015cia}
\eq{\tilde{J}^I=J^J(M^{-1})_J{}^I~,~\widetilde{\nabla}_I=M_I{}^J\nabla_J	~,~
	M_I{}^J=(L^{-1})_I{}^MR_M{}^J~,~\eta_{IJ}=M_I{}^KM_J{}^L\eta_{KL} \label{34}~~~.}

The affine extension of the algebras \bref{nab} is given 
by generalization $\nabla_I\to \dd_I(\sigma)$ and  $\widetilde{\nabla}_I\to \widetilde{\dd}_I(\sigma)$ .
The worldsheet indices of the currents are denoted as $m=(\tau,\sigma)=(0,1)$ for currents
$J_m{}^I$ and  $\tilde{J}_m{}^I$ with $dZ^M=d\sigma^m\partial_m Z^M$. 
The affine covariant derivative and the symmetry generator are given by
\eq{{\renewcommand{\arraystretch}{1.6}
		\begin{array}{lccl}	
			{\renewcommand{\arraystretch}{0.9}\begin{array}{c}\rm{Covariant~derivative}\\(\rm{Selfdual~current})
\end{array}		}	
			:&	\dd_I&=&\nabla_I+J_1^JN_{I}{}^K\eta_{KJ}=\nabla_I+J_1^J(B_{JI}+\eta_{JI})\\
			\rm{Symmetry~generator}:&		\widetilde{\dd}_I&=&\widetilde{\nabla}_I-\tilde{J}_1^JM_J{}^LM_{I}{}^KN_{LK}
			=M_I{}^J\hat{\dd}_J\\
		\rm{Anti}{\mathchar`-}\rm{selfdual~current}	&\hat{\dd}_J&\equiv& \nabla_J+J_1^K(B_{KJ}-\eta_{KJ})		\end{array}\label{epmzwe2}}}
We consider cases where $B_{IJ}$ is constant and determined by the dilatation operator as \cite{Hatsuda:2015cia}
\eq{B_{IJ}=\f12N_{[I|}{}^{K}\eta_{K|J)}=\f12(n_J-n_I)\eta_{IJ}}
where the dilatation operator $\hat{N}$ gives the canonical dimension of the generator $G_I$
with the following normalization
\eq{[\hat{N},G_I]=iN_I{}^JG_J=in_IG_I~~,~~(n_I+n_J)\eta_{IJ}=2\eta_{IJ}~~~.}
They satisfy the following affine Lie algebra
\bea
{\renewcommand{\arraystretch}{1.6}
	\left\{
	\begin{array}{ccl}
		\lbrack \dd_I(1),\dd_J(2)\}&=&-if_{IJ}{}^K\dd_K\delta(2-1)-2i\eta_{IJ}\partial_\sigma\delta(2-1)\\
		\lbrack \widetilde{\dd}_I(1),\widetilde{\dd}_J(2)\}&=&~if_{IJ}{}^K\widetilde{\dd}_K\delta(2-1)+2i\eta_{IJ}\partial_\sigma\delta(2-1)\\
		\lbrack \dd_I(1),\widetilde{\dd}_J(2)\}&=&0 
	\end{array}\right. }
\eea
where the worldsheet  $\sigma$ coordinates $\sigma_1, \sigma_2$ are denoted as 
$1, 2$ and $\partial_\sigma\delta(2-1)=\frac{\partial}{\partial \sigma_2}\delta(\sigma_2-\sigma_1)$.

In general the Hamiltonian is written as bilinears of the covariant derivatives  $\dd_I$,
while the global symmetry charge is given by the integral of the symmetry generator $\tilde{\dd}_I$
so that the Hamiltonian is invariant under the global symmetry.
The 2-dimensional operators  $\dd_I$ and $\hat{\dd}_I$, which is  O(D,D) transformed   $\tilde{\dd}_I$ in \bref{epmzwe2},  in manifestly T-duality formulation  are 
seldfual and anti-selfdual  respectively.
Some of the symmetry generators are set to be zero
as the dimensional reduction constraints.

\subsection{Virasoro and  selfduality constraints}

We double the algebra \bref{GGfG} as the direct product of two copies of the algebra with the opposite sign of the structure constant:
\eq{\renewcommand{\arraystretch}{1.6}
	\begin{array}{ccl}
		G_I&\to&G_I=(G_{\bar{I}}~,~G_{\underline{I}})\\
		Z^M&\to&Z^M=(Z^{\overline{M}}~,~Z^{\underline{M}})\\
		f_{IJ}{}^K&\to&f_{IJ}{}^K=(f_{\bar{I}\bar{J}}{}^{\overline{K}}~,~ f_{\underline{IJ}}{}^{\underline{K}}=
		-f_{\bar{I}\bar{J}}{}^{\overline{K}})\\
		\eta_{IJ}&\to&\eta_{IJ}=(\eta_{\bar{I}\bar{J}}~,~\eta_{\underline{I}\underline{J}}=-\eta_{\bar{I}\bar{J}})~,~
		\hat{\eta}_{IJ}=(\eta_{\bar{I}\bar{J}}~,~\eta_{\underline{I}\underline{J}}=\eta_{\bar{I}\bar{J}})
\end{array}}
where two kinds of metrics are written  in matrix notarion as
\eq{ \eta_{IJ}=	\left(\begin{array}{cc}  \eta_{\bar{I}\bar{J}}&0\\0&-\eta_{\bar{I}\bar{J}}\end{array}	\right)
	~,~	\hat{\eta}_{IJ}=	\left(\begin{array}{cc}  \eta_{\bar{I}\bar{J}}&0\\0&\eta_{\bar{I}\bar{J}}\end{array}	\right)~~~.}
The bilinears of currents \bref{epmzwe2} contracted with these metrics are the Virasoro generators
\bea
{\renewcommand{\arraystretch}{1.6}
	\left\{
	\begin{array}{ccl}
		h_\sigma&=&\f14	{\dd}{}_I{{\eta}}^{IJ}{\dd}{}_J	= \f14 \left( ({\dd}{}_{\bar{I}})^2-(\dd_{\underline{I}})^2\right)\\
		h_\tau&=&\f14{\dd}{}_I\hat{\eta}^{IJ}{\dd}{}_J
		=\f14 	\left(	(\dd_{\bar{I}})^2+(\dd_{\underline{I}})^2\right)	\\	
		\tilde{h}_\sigma&=&\f14 \widetilde{\dd}_I{{\eta}}^{IJ}\widetilde{\dd}_J=\f14\left((\widetilde{\dd}_{\bar{I}})^2	-(\widetilde{\dd}_{\underline{I}})^2\right)\\
		&=&\f14\hat{\dd}_I{{\eta}}^{IJ}\hat{\dd}_J
		=\f14\left((\hat{\dd}_{\bar{I}})^2	-(\hat{\dd}_{\underline{I}})^2\right)	\\
		\tilde{h}_\tau&=&\f14 \widetilde{\dd}_I \hat{\eta}^{IJ} \widetilde{\dd}_J=\f14\left((\widetilde{\dd}_{\bar{I}})^2		+(\widetilde{\dd}_{\underline{I}})^2\right)
	\end{array}\right.}
\eea
where $\widetilde{\dd}_I =M_I{}^J\hat{\dd}_J$ in \bref{epmzwe2} is used in the fourth line.
The Virasoro algebras are given by
\bea
&&{\renewcommand{\arraystretch}{1.6}
	\left\{\begin{array}{ccl}
		\lbrack h_\sigma(1),h_\sigma(2)]&=&-i\left( h_\sigma(1)+h_\sigma(2)\right)\partial_\sigma\delta(2-1)\\
		\lbrack h_\sigma(1), h_\tau(2)]&=&-i\left( h_\tau(1)+ h_\tau(2)\right)\partial_\sigma\delta(2-1)\\
		\lbrack h_\tau(1), h_\tau(2)]&=&-i\left( h_\sigma(1)+ h_\sigma(2)\right)\partial_\sigma\delta(2-1)
	\end{array}\right.}\label{Virasorohhh}\\
&&{\renewcommand{\arraystretch}{1.6}
	\left\{\begin{array}{ccl}
		\lbrack \tilde{h}_\sigma(1),\tilde{h}_\sigma(2)]&=&i\left( \tilde{h}_\sigma(1)+\tilde{h}_\sigma(2)\right)\partial_\sigma\delta(2-1)\\
		\lbrack \tilde{h}_\sigma(1), \tilde{h}_\tau(2)]&=&i\left( \tilde{h}_\tau(1)+ \tilde{h}_\tau(2)\right)\partial_\sigma\delta(2-1)\\
		\lbrack \tilde{h}_\tau(1), \tilde{h}_\tau(2)]&=&i\left( \tilde{h}_\sigma(1)+ \tilde{h}_\sigma(2)\right)\partial_\sigma\delta(2-1)
	\end{array}\right.}\\
&&{\renewcommand{\arraystretch}{1.6}
	\left\{\begin{array}{ccl}
		\lbrack {h}_m(1),\tilde{h}_n(2)]&=&0
	\end{array}\right.} ~~,~~m,n=(\tau,\sigma)
\eea

Derivative operators act on fields $\Phi(Z)$ as $Z^M$ derivatives
\eq{	\lbrack \dd_I,\Phi]=\lbrack \hat{\dd}_I,\Phi]=\frac{1}{i}(R^{-1})_I{}^M\partial_M \Phi(Z)~~~.  }
The $\sigma$ derivative is defined by the commutator with both selfdual and anti-selfdual Virasoro operators,
so that the $\sigma$ derivative in canonical formalism coincides with the usual chain rule derivative
\eq{\partial_\sigma \Phi(Z)= i	\lbrack \displaystyle \int (h_\sigma-\tilde{h}_\sigma),\Phi]
	=\f12(\dd_I -\hat{\dd}_I)\eta^{IJ} (R^{-1})_I{}^M\partial_M\Phi
	=\partial_\sigma Z^M\partial_{M}\Phi}
Therefore we take a set of constraints for a string system in the doubled space;
the Virasoro constraints and the selfduality constraints as
\eq{{\renewcommand{\arraystretch}{1.3}
	{\renewcommand{\arraystretch}{0.9}\begin{array}{c}\rm{Virasoro}\\ \rm{constraints}
		\end{array}	}~	\left\{\begin{array}{ccl}
			{\mathcal H}_\sigma&=&	h_\sigma-\tilde{h}_\sigma=0\\
			{\mathcal H}_\tau&=&	h_\tau+\tilde{h}_\tau=0
		\end{array}\right.~~,~~ {\renewcommand{\arraystretch}{0.9}\begin{array}{c}\rm{Selfduality}\\ \rm{constraints}
	\end{array}	}
~\left\{\begin{array}{ccl}
			\tilde{h}_\sigma&=&0\\
			\tilde{h}_\tau&=&0
		\end{array}\right.}}
$	{\mathcal H}_\sigma$ and $	{\mathcal H}_\tau$ satisfy the same Virasoro algebra in \bref{Virasorohhh}.

\section{T-dual superstring Lagrangians}

\subsection{Superalgebras and currents}

In the manifestly T-duality formulation a superstring in a flat space is governed by 
the doubled nondegenerate superalgebra generated by $G_{\cal M}$.
The doubled indices for the left and right sectors are denoted by ${\cal M}=(\overline{\cal M},\underline{\cal M})$.
The nondegenerate supergenerators are denoted as $G_{\cal M}=(d_\mu,~P_M,~\omega^\mu)$
$=(d_{\bar{\mu}},~P_{\overline{M}},~\omega^{\bar{\mu}};~d_{\underline{\mu}},~P_{\underline{M}},~\omega^{\underline{\mu}})$.
The algebra is given by
\eq{
	{\renewcommand{\arraystretch}{1.6}
		\begin{array}{llcl}
			\rm{Left}:&\{d_{\bar{\mu}},d_{\bar{\nu}}  \}=2P_{\overline{M}}\gamma^{\overline{M}}{}_{\bar{\mu}\bar{\nu}}&,&
			\lbrack	 d_{\bar{\mu}},P_{\overline{M}}]=2(\gamma_{\overline{M}}\omega)_{\bar{\mu}}		\\
			\rm{Right}:&\{d_{\underline{\mu}},d_{\underline{\nu}}  \}=-2P_{\underline{M}}\gamma^{\underline{M}}{}_{\underline{\mu}\underline{\nu}}&,&
			\lbrack	 d_{\underline{\mu}},P_{\underline{M}}]=-2(\gamma_{\underline{M}}\omega)_{\underline{\mu}}		
\end{array} }}
The nondegenerate metric $\eta_{MN}$ and  $B_{MN}$ are given as
\bea
&&\eta_{\cal MN}=
{\renewcommand{\arraystretch}{1.6}
	\left(
	\begin{array}{c|c}\eta_{\overline{\cal M}\overline{\cal N}}&0\\\hline 0&
		-\eta_{\overline{\cal M}\overline{\cal N}}
	\end{array}
	\right)}
~,~\eta_{\overline{\cal M}\overline{\cal N}}
=\begin{array}{c}d_{\bar{\mu}}\\P_{\overline{M}}\\\omega^{\bar{\mu}}
\end{array}
\left(\begin{array}{ccc}&&{\bf 1}\\&{\bf 1}&\\-{\bf 1}&&\end{array}
\right)~
\\
&&B_{\cal MN}=	{\renewcommand{\arraystretch}{1.6}\left(
	\begin{array}{c|c}B_{\overline{\cal M}\overline{\cal N}}&0\\\hline 0&
		-B_{\overline{\cal M}\overline{\cal N}}
	\end{array}
	\right)}
~,~B_{\overline{\cal M}\overline{\cal N}}
=\begin{array}{c}d_{\bar{\mu}}\\P_{\overline{M}}\\\omega^{\bar{\mu}}
\end{array}
\left(\begin{array}{ccc}&&\frac{\bf 1}{2}\\&{ 0}&\\\frac{\bf 1}{2}&&\end{array}
\right)
\eea
where blank spaces are zeros.

For a group element $g$ the left-invariant and the right-invariant currents are denoted by
\eq{	{\renewcommand{\arraystretch}{1.6}
		\begin{array}{ccl}
			\frac{1}{i}g^{-1}dg&=&J^{\cal M} G_{\cal M}=
			J^{\bar{\mu}} d_{\bar{\mu}}
			+J^{\overline{M}} P_{\overline{M}}
			+J_{\bar{\mu}} \omega^{\bar{\mu}}
			+J^{\underline{\mu}} d_{\underline{\mu}}
			+J^{\underline{M}}  P_{\underline{M}}
			+J_{\underline{\mu}}\omega^{\underline{\mu}}\\
			\frac{1}{i}dgg^{-1}&=&\tilde{J}^{\cal M} G_{\cal M}=
			\tilde{J}^{\bar{\mu}} d_{\bar{\mu}}
			+\tilde{J}^{\overline{M}} P_{\overline{M}}
			+\tilde{J}_{\bar{\mu}} \omega^{\bar{\mu}}
			+\tilde{J}^{\underline{\mu}} d_{\underline{\mu}}
			+\tilde{J}^{\underline{M}}  P_{\underline{M}}
			+\tilde{J}_{\underline{\mu}}\omega^{\underline{\mu}}\end{array}}}
The Maurer-Cartan equations for the left/right one form currents are given as
\bea
&&\rm{Left}{\mathchar`-}\rm{invariant~currents}\nn\\
&&\rm{Left}:{\renewcommand{\arraystretch}{1.3}
	\left\{
	\begin{array}{ccl}
		dJ^{\bar{\mu}}&=&0\\	
		dJ^{\overline{M}}&=&iJ^{\bar{\mu}}\wedge J^{\bar{\nu}}\gamma^{\overline{M}}{}_{\bar{\nu}\bar{\mu}}\\
		dJ_{\bar{\mu}}&=&-2i J^{\bar{\nu}}\wedge J^{\overline{N}}\gamma_{\overline{N}}{}_{\bar{\nu}\bar{\mu}}	
	\end{array}\right.}~~,~~
\rm{Right}:{\renewcommand{\arraystretch}{1.3}
	\left\{
	\begin{array}{ccl}
		dJ^{\underline{\mu}}&=&0\\
		dJ^{\underline{M}}&=&-iJ^{\underline{\mu}}\wedge J^{\underline{\nu}}\gamma^{\underline{M}}{}_{\underline{\nu}\bar{\mu}}\\
		dJ_{\underline{\mu}}&=&2iJ^{\underline{\nu}}\wedge J^{\underline{N}}\gamma_{\underline{N}}{}_{\underline{\nu\mu}}	
	\end{array}\right.}	\label{MCLeft}\\
&&\rm{Right}{\mathchar`-}\rm{invariant~currents}\nn\\
&&\rm{Left}:{\renewcommand{\arraystretch}{1.3}
	\left\{
	\begin{array}{ccl}
		d\tilde{J}^{\bar{\mu}}&=&0\\	
		d\tilde{J}^{\overline{M}}&=&-i\tilde{J}^{\bar{\mu}}\wedge \tilde{J}^{\bar{\nu}}\gamma^{\overline{M}}{}_{\bar{\nu}\bar{\mu}}\\
		d\tilde{J}_{\bar{\mu}}&=&2i \tilde{J}^{\bar{\nu}}\wedge \tilde{J}^{\overline{N}}\gamma_{\overline{N}}{}_{\bar{\nu}\bar{\mu}}	
	\end{array}\right.}~~,~~
\rm{Right}:{\renewcommand{\arraystretch}{1.3}
	\left\{
	\begin{array}{ccl}
		d\tilde{J}^{\underline{\mu}}&=&0\\
		d\tilde{J}^{\underline{M}}&=&i\tilde{J}^{\underline{\mu}}\wedge \tilde{J}^{\underline{\nu}}\gamma^{\underline{M}}{}_{\underline{\nu}\bar{\mu}}\\
		d\tilde{J}_{\underline{\mu}}&=&-2iJ^{\underline{\nu}}\wedge \tilde{J}^{\underline{N}}\gamma_{\underline{N}}{}_{\underline{\nu\mu}}	
	\end{array}\right.}	
\eea

A  group element $g(Z^{\cal M})$ is parametrized with $Z^{\cal M}=(\theta^{\bar{\mu}}, X^{\overline{M}},\varphi_{\bar{\mu}}; ~
\theta^{\underline{\mu}}, X^{\underline{X}},\varphi_{\underline{\mu}})$ by
\bea
&&g(Z^{\cal M})=g(Z^{\overline{\cal M}})g(Z^{\underline{\cal M}})\nn\\
&&	{\renewcommand{\arraystretch}{1.6}
	\left\{\begin{array}{ccl}
		g(Z^{\overline{\cal M}})&=&\rm{exp}(i\varphi_{\bar{\mu}}\omega^{\bar{\mu}})
		\rm{exp}(iX^{\overline{M}}P_{\overline{M}})\rm{exp}(i\theta^{\bar{\mu}}d_{\bar{\mu}})
		\\
		g(Z^{\underline{\cal M}})&=&\rm{exp}(i\varphi_{\underline{\mu}}\omega^{\underline{\mu}})
		\rm{exp}(iX^{\underline{M}}P_{\underline{M}})\rm{exp}(i\theta^{\underline{\mu}}d_{\underline{\mu}})
	\end{array}\right.}
\eea 
The left/right-invariant currents  are given as
\bea
&&\rm{Left}{\mathchar`-}\rm{invariant~currents}\nn\\
&&\rm{Left}:~~{\renewcommand{\arraystretch}{1.3}
	\left\{\begin{array}{ccl}
		J^{\bar{\mu}}&=&d\theta^{\bar{\mu}}\\
		J^{\overline{M}}&=&dX^{\overline{M}}-i\theta\gamma^{\overline{M}}d\theta\\
		J_{\bar{\mu}}&=&d\varphi_{\bar{\mu}}-2i(dX^{\overline{M}}-\frac{i}{3}\theta\gamma^{\overline{M}}d\theta)(\theta\gamma_{\overline{M}})_{\bar{\mu}}		
	\end{array}\right.  }\nn\\
&&\rm{Right}:~{\renewcommand{\arraystretch}{1.3}   
	\left\{\begin{array}{ccl}
		J^{\underline{\mu}}&=&d\theta^{\underline{\mu}}\\
		J^{\underline{M}}&=&dX^{\underline{M}}+i\theta\gamma^{\underline{M}}d\theta\\
		J_{\underline{\mu}}&=&d\varphi_{\underline{\mu}}+2i(dX^{\underline{M}}+\frac{i}{3}\theta\gamma^{\underline{M}}d\theta)(\theta\gamma_{\underline{M}})_{\underline{\mu}}		
	\end{array}\right.~}\nn\\
&&\rm{Right}{\mathchar`-}\rm{invariant~currents}\nn\\
&&\rm{Left}:~~{\renewcommand{\arraystretch}{1.3}
	\left\{\begin{array}{ccl}
		\tilde{J}^{\bar{\mu}}&=&d\theta^{\bar{\mu}}\\
		\tilde{J}^{\overline{M}}&=&dX^{\overline{M}}+i(\theta\gamma^{\overline{M}}d\theta)\\
		\tilde{J}_{\bar{\mu}}&=&d\varphi_{\bar{\mu}}-2i(dX^{\overline{M}}-\frac{i}{3}\theta\gamma^{\overline{M}}d\theta)(\theta\gamma_{\overline{M}})_{\bar{\mu}}		
	\end{array}\right.}\nn\\
&&\rm{Right}:~{\renewcommand{\arraystretch}{1.3}
	\left\{\begin{array}{ccl}
		\tilde{J}^{\underline{\mu}}&=&d\theta^{\underline{\mu}}\\
		\tilde{J}^{\underline{M}}&=&dX^{\underline{M}}+i\theta\gamma^{\underline{M}}d\theta\\
		\tilde{J}_{\underline{\mu}}&=&d\varphi_{\underline{\mu}}+2i(dX^{\underline{M}}+\frac{i}{3}\theta\gamma^{\underline{M}}d\theta)(\theta\gamma_{\underline{M}})_{\underline{\mu}}		
	\end{array}\right.}
\eea 
The left/right-invariant derivatives are given as
\bea
&&\rm{Left}{\mathchar`-}\rm{invariant~currents}\nn\\
&&\rm{Left}:~~{\renewcommand{\arraystretch}{1.3}
	\left\{\begin{array}{ccl}
		\nabla_{\bar{\mu}}&=&-i\partial_{\bar{\mu}}
		+(\theta \gamma^{\overline{M}})_{\bar{\mu}}\partial_{\overline{M}}
		+i\frac{4}{3}(\theta\gamma^{\overline{M}})_{\bar{\mu}} (\theta\gamma_{\overline{M}})_{\bar{\nu}}\partial^{\bar{\nu}}	\\
		\nabla_{\overline{M}}&=&-i\partial_{\overline{M}}+2(\theta\gamma^{\overline{M}})_{\bar{\mu}}\partial^{\bar{\nu}}\\
		\nabla^{\bar{\mu}}&=&-i\partial^{\bar{\mu}}\end{array}\right.  }\\
&&\rm{Right}:~{\renewcommand{\arraystretch}{1.3}   
	\left\{\begin{array}{ccl}
		\nabla_{\underline{\mu}}&=&-i\partial_{\underline{\mu}}
		-(\theta \gamma^{\underline{M}})_{\underline{\mu}}\partial_{\underline{M}}
		+i\frac{4}{3}(\theta\gamma^{\underline{M}})_{\underline{\mu}} (\theta\gamma_{\underline{M}})_{\underline{\nu}}\partial^{\underline{\nu}}	\\
		\nabla_{\underline{M}}&=&-i\partial_{\underline{M}}-2(\theta\gamma^{\underline{M}})_{\underline{\nu}}\partial^{\underline{\nu}}\\
		\nabla^{\underline{\mu}}&=&-i\partial^{\underline{\mu}}\end{array}\right.  }\\
&&\rm{Right}{\mathchar`-}\rm{invariant~currents}\nn\\
&&\rm{Left}:~~{\renewcommand{\arraystretch}{1.3}
	\left\{\begin{array}{ccl}
		\widetilde{\nabla}_{\bar{\mu}}&=&-i\partial_{\bar{\mu}}
		-(\theta \gamma^{\overline{M}})_{\bar{\mu}}\partial_{\overline{M}}
		+2X_{\overline{M}}\gamma^{\overline{M}}{}_{\bar{\mu}\bar{\nu}}\partial^{\bar{\nu}} -i\frac{2}{3}(\theta\gamma^{\overline{M}})_{\bar{\mu}}(\theta\gamma_{\overline{M}})_{\bar{\nu}}
		\partial^{\bar{\nu}}	\\
		\widetilde{\nabla}_{\overline{M}}&=&-i\partial_{\overline{M}}\\
		\widetilde{\nabla}^{\bar{\mu}}&=&-i\partial^{\bar{\mu}}\end{array}\right.}\\
&&\rm{Right}:~{\renewcommand{\arraystretch}{1.3}
	\left\{\begin{array}{ccl}
		\widetilde{\nabla}_{\underline{\mu}}&=&-i\partial_{\underline{\mu}}
		+(\theta \gamma^{\underline{M}})_{\underline{\mu}}\partial_{\underline{M}}
		-2X_{\underline{M}}\gamma^{\underline{M}}{}_{\underline{\mu}\underline{\nu}}\partial^{\underline{\nu}} -i\frac{2}{3}(\theta\gamma^{\underline{M}})_{\underline{\mu}}(\theta\gamma_{\underline{M}})_{\underline{\nu}}
		\partial^{\underline{\nu}}	\\
		\widetilde{\nabla}_{\underline{M}}&=&-i\partial_{\underline{M}}\\
		\widetilde{\nabla}^{\underline{\mu}}&=&-i\partial^{\underline{\mu}}\end{array}\right.}
\eea 
The covariant derivatives $\dd_{\cal M}$  and the anti-selfdual currents $\hat{\dd}_{\cal M}$ which are proportional to  the symmetry generators $\widetilde{\dd}_{\cal M}=M_{\cal M}{}^{\cal N}\hat{\dd}_{\cal N}$ as in \bref{nab} and \bref{34} are the followings:
\bea
&&\rm{Selfdual~current},~~ \dd_{\cal M}=(D_\mu,~{\cal P}_M~,\Omega^{\mu}) \nn\\
&&\rm{Left}:~{\renewcommand{\arraystretch}{1.3}
	\left\{\begin{array}{ccl}
		D_{\bar{\mu}}&=&\nabla_{\bar{\mu}}-\f12J_1{}_{\bar{\mu}}\\
		{\cal P}_{\overline{M}}&=&\nabla_{\overline{M}}+J_1{}_{\overline{M}}\\
		\Omega^{\bar{\mu}}&=&\nabla^{\bar{\mu}}+\frac{3}{2}J_1{}^{\bar{\mu}}\end{array}\right.
	~,~\rm{Right}:~
	\left\{\begin{array}{ccl}
		D_{\underline{\mu}}&=&\nabla_{\underline{\mu}}+\f12J_1{}_{\underline{\mu}}\\
		{\cal P}_{\underline{M}}&=&\nabla_{\underline{M}}-J_1{}_{\underline{M}}\\
		\Omega^{\underline{\mu}}&=&\nabla^{\underline{\mu}}-\frac{3}{2}J_1{}^{\underline{\mu}}\end{array}\right.}\\
&&\rm{Anti}{\mathchar`-}\rm{selfdual~currents},~~ \hat{\dd}_{\cal M}=(\hat{D}_\mu,~\hat{\cal P}_M~,\hat{\Omega}^{\mu}) \nn\\
&&\rm{Left}:~{\renewcommand{\arraystretch}{1.3}
	\left\{\begin{array}{ccl}
		\hat{D}_{\bar{\mu}}&=&\nabla_{\bar{\mu}}+\frac{3}{2}J_1{}_{\bar{\mu}}\\
		\hat{\cal P}_{\overline{M}}&=&\nabla_{\overline{M}}-J_1{}_{\overline{M}}\\
		\hat{\Omega}^{\bar{\mu}}&=&\nabla^{\bar{\mu}}-\f12J_1{}^{\bar{\mu}}\end{array}\right.
	~,~\rm{Right}:~
	\left\{\begin{array}{ccl}
		\hat{\D}_{\underline{\mu}}&=&\nabla_{\underline{\mu}}-\frac{3}{2}J_1{}_{\underline{\mu}}\\
		\hat{\cal P}_{\underline{M}}&=&\nabla_{\underline{M}}+J_1{}_{\underline{M}}\\
		\hat{\Omega}^{\underline{\mu}}&=&\nabla^{\underline{\mu}}+\f12J_1{}^{\underline{\mu}}\end{array}\right.}
\eea

\subsection{Superstring Lagrangians}

We propose the superstring Lagrangian with manifestly T-duality 
 in Hamilton form 
\bea{\renewcommand{\arraystretch}{1.6}\begin{array}{ccl}
		L&=&\partial_\tau{Z}^{\cal M}P_{\cal M}-H\\
		H&=&g_-\f14\dd_{\overline{\cal M}}\eta^{\overline{\cal M}\overline{\cal N}} \dd_{\overline{\cal N}}
		+ g_+\f14\dd_{\underline{\cal M}}\eta^{\overline{\cal M}\overline{\cal N}} \dd_{\underline{\cal N}}\\
		&&
		+(g_++\lambda_+)\f14\widetilde{\dd}_{\overline{\cal M}}\eta^{\overline{\cal M}\overline{\cal N}} \widetilde{\dd}_{\overline{\cal N}}
		+(g_-+\lambda_-)\f14\widetilde{\dd}_{\underline{\cal M}}\eta^{\overline{\cal M}\overline{\cal N}}\widetilde{\dd}_{\underline{\cal N}}\\
		&& +\bar{\Lambda}\widetilde{\Omega}^{\bar{\mu}} +\underline{\Lambda}\widetilde{\Omega}^{\underline{\mu}}
\end{array}}
\eea
where the set of first class constraints are the Virasoro constraints,
the selfduality constraints and the dimensional reduction constraints of auxiliary fermions
\eq{
	{\renewcommand{\arraystretch}{1.6}
		\left\{\begin{array}{ccl}
			{\mathcal H}_\tau &=& \displaystyle\f14(\dd_{{\cal M}}\hat{\eta}^{{\cal M}{\cal N}} \dd_{{\cal N}}
			-\widetilde{\dd}_{{\cal M}}\hat{\eta}^{{\cal M}{\cal N}}\widetilde{\dd}_{{\cal N}})=0 \\
			&=&\displaystyle\f14(\dd_{\overline{\cal M}}\eta^{\overline{\cal M}\overline{\cal N}}\dd_{\overline{\cal N}} +
			\dd_{\underline{\cal M}}\eta^{\overline{\cal M}\overline{\cal N}} \dd_{\underline{\cal N}}
			+\widetilde{\dd}_{\overline{\cal M}}\eta^{\overline{\cal M}\overline{\cal N}}\widetilde{\dd}_{\overline{\cal N}} +
			\widetilde{\dd}_{\underline{\cal M}}\eta^{\overline{\cal M}\overline{\cal N}}\widetilde{\dd}_{\underline{\cal N}} )\\
			{\mathcal H}_\sigma &=& \displaystyle\f14
			(\dd_{{\cal M}}{\eta}^{{\cal M}{\cal N}} \dd_{{\cal N}}
			-\hat{\dd}_{{\cal M}}\eta^{{\cal M}{\cal N}}\hat{\dd}_{{\cal N}} )=0\\
			&=&\displaystyle\f14(\dd_{\overline{\cal M}}\eta^{\overline{\cal M}\overline{\cal N}}\dd_{\overline{\cal N}} -
			\dd_{\underline{\cal M}}\eta^{\overline{\cal M}\overline{\cal N}} \dd_{\underline{\cal N}}
			-\hat{\dd}_{\overline{\cal M}}\eta^{\overline{\cal M}\overline{\cal N}}\hat{\dd}_{\overline{\cal N}} +
			\hat{\dd}_{\underline{\cal M}}\eta^{\overline{\cal M}\overline{\cal N}}\hat{\dd}_{\underline{\cal N}} )\\
			\bar{\chi}&=&\widetilde{\dd}_{\overline{\cal M}}\eta^{\overline{\cal M}\overline{\cal N}}\widetilde{\dd}_{\overline{\cal N}}=0\\
			\underline{\chi}&=&\widetilde{\dd}_{\underline{\cal M}}\eta^{\overline{\cal M}\overline{\cal N}}\widetilde{\dd}_{\underline{\cal N}}=0\\
			\widetilde{\Omega}^{\bar{\mu}}&=&\hat{\Omega}^{\bar{\mu}}=0\\
			\widetilde{\Omega}^{\underline{\mu}}&=&\hat{\Omega}^{\bar{\mu}}=0
		\end{array}\right.}}
In the last two lines 
 $M^{\mu}{}^{\nu}=0=M^{\mu}{}_{N}$ and $M^{\mu}{}_{\nu}=\delta_\mu^\nu$ of 
 $M_{\mathcal M}{}^{\mathcal N}=(L^{-1}{}R)_{\mathcal M}{}^{\mathcal N}$
 are used.
The Hamiltonian becomes
\bea
H&=&g_-\displaystyle(\f14{\cal P}_{\overline{M}}{}^2+\f12\Omega^{\bar{\mu}}D_{\bar{\mu}})
+g_+\displaystyle(\f14{\cal P}_{\underline{M}}{}^2
+\f12\Omega^{\underline{\mu}}D_{\underline{\mu}})
+(g_++\lambda_+)\f14\hat{\cal P}_{\overline{M}}{}^2
+(g_-+\lambda_-)\f14\hat{\cal P}_{\underline{M}}{}^2
\nn\\
&&+\bar{\Lambda}\widetilde{\Omega}^{\bar{\mu}} +\underline{\Lambda}\widetilde{\Omega}^{\underline{\mu}}\label{Hamsust}
\eea
The Legendre tranfsormation brings to the following Lagrangian
\bea
&&L_1=L_{1:\rm kin}+L_{1:\rm WZ}\nn\\
&&	{\renewcommand{\arraystretch}{1.6}\left\{
	\begin{array}{ccl}
		{L}_{1:\rm kin}&=&\displaystyle\frac{1}{\bar{e}}~(\bar{e}_+{}^m{J}_m^{\overline{M}})(\bar{e}_-{}^n J_n^{\overline{N}}) \eta_{\overline{M}\overline{N}}+
		\displaystyle\frac{1}{\underline{e}}~(\underline{e}_+{}^mJ_m{}^{\underline{M}}) (\underline{e}_-{}^nJ_n{}^{\underline{N}})\eta_{\overline{M}\overline{N}} \\
		{L}_{1:\rm WZ}&=&
		\displaystyle\frac{k}{2}		(		{J}_{[0}{}^{\bar{\mu}} {J}_{1]}{}_{\bar{\mu}}
		-		{J}_{[0}{}^{\underline{\mu}} {J}_{1]}{}_{\underline{\mu}} 		)		\\
		&=&\displaystyle\frac{k}{2}\left[
		\frac{1}{\bar{e}}~
		(\bar{e}_{[+|}{}^m J_m{}^{\bar{\mu}})(\bar{e}_{|-]}{}^n J_n{}_{\bar{\mu}})
		-\frac{1}{\underline{e}}~
		(\underline{e}_{[+|}{}^mJ_m{}^{\underline{\mu}}) (\underline{e}_{|-]}{}^nJ_n{}_{\underline{\mu}})
		\right]
	\end{array}\right.\label{doublezweibeins}}
\eea
This is one of the main results, the superstring Lagrangian with manifest T-duality with
double zweibeins 
where $\bar{e}_a{}^m$  and  $\underline{e}_a{}^m$ are zweibeins  in \bref{ebareul}.
The worldsheet contraction $\epsilon_{ab}=\frac{1}{e}e_a{}^me_b{}^n\epsilon_{mn}
=\frac{1}{\bar{e}}\bar{e}_a{}^m\bar{e}_b{}^n\epsilon_{mn}=\frac{1}{\underline{e}}\underline{e}_a{}^m\underline{e}_b{}^n\epsilon_{mn}
$ with $\epsilon_{01}=-\epsilon_{10}=1$ is used.
The Hamiltonian \bref{Hamsust} gives $k=1$, and we generalize the constant $k=\pm 1$ as usual.

Doubling zweibein makes computation of the $\kappa$-symmetry transformation easier,
since each left and right sector Lagrangians are $\kappa$-symmetry invariant
separately.
The $\kappa$-symmetry transformation $\delta_\kappa g$ is denoted by $\Delta^{\cal M}$
and the vector components $\Delta^M$ vanish as
\eq{g^{-1}\delta_\kappa g=i\Delta^{\cal M}G_{\cal M}~~,~~\Delta^M=(\Delta^{\overline{M}},~\Delta^{\underline{M}})=0~~~.}
Since the left and right sectors are completely independent,
we just use $e_a{}^m$ as the zweibeins 
instead of $\bar{e}_a{}^m$ and $\underline{e}_a{}^m$
to perform the $\kappa$-invariance computation
only in this paragraph.
The $\kappa$-symmetry transformation of zwiebein $e_m{}^a$ is denoted as
\eq{\delta_\kappa (\frac{1}{\sqrt{e}}e_a{}^m)~\sqrt{e}e_m{}^b=\Delta_a{}^b~~~}
which is traceless, $\Delta_a{}^a=0$.
The worldsheet currents with the zwiebein is denoted as $J_a{}^{\cal M}=\frac{1}{\sqrt{e}}e_a{}^mJ_m{}^{\cal M}$.
Under the $\kappa$-symmetry transformation the currents are transformed as
\eq{\delta_\kappa J_m{}^{\cal M}=\partial_m \Delta^{\cal M}+J_m{}^{\cal L}\Delta^{\cal N}f_{\cal NL}{}^{\cal M}~~,~~
	\delta_\kappa J_a{}^{\cal M}=\Delta_a{}^bJ_b{}^{\cal M}
	+\partial_a \Delta^{\cal M}
	+J_a{}^{\cal L}\Delta^{\cal N}f_{\cal NL}{}^{\cal M}}
with $\partial_a=\frac{1}{\sqrt{e}}e_a{}^m\partial_m$.
It is convenient to write down the $\kappa$-symmetry transformation of the left and right currents
\eq{	{\renewcommand{\arraystretch}{1.6}
		\rm{Left}:~	\left\{
		\begin{array}{ccl}
			\delta_\kappa J_a{}^{\bar{\mu}}&=&\Delta_a{}^bJ_b{}^{\bar{\mu}}+\partial_a\Delta^{\bar{\mu} }\\
			\delta_\kappa J_a{}^{\overline{M}}&=&\Delta_a{}^bJ_b{}^{\overline{M}}
			+{2}{i}\Delta^{\bar{\mu}}J_a{}^{\bar{\nu}}
			\gamma^{\overline{M}}{}_{\bar{\mu}\bar{\nu}}\\
			\delta_\kappa J_a{}_{\bar{\mu}}&=&
			\Delta_a{}^bJ_b{}_{\bar{\mu}}+\partial_a\Delta_{\bar{\mu} }
			-{2}{i}\Delta^{\bar{\nu}}J_a{}^{\overline{M}}
			\gamma_{\overline{M}}{}_{\bar{\mu}\bar{\nu}}	
		\end{array}\right.}}
\eq{	{\renewcommand{\arraystretch}{1.6}
		\rm{Right}:~	\left\{
		\begin{array}{ccl}
			\delta_\kappa J_a{}^{\underline{\mu}}&=&\Delta_a{}^bJ_b{}^{\underline{\mu}}+\partial_a\Delta^{\underline{\mu} }\\
			\delta_\kappa J_a{}^{\underline{M}}&=&\Delta_a{}^bJ_b{}^{\underline{M}}
			-{2}{i}\Delta^{\underline{\mu}}J_a{}^{\underline{\nu}}
			\gamma^{\underline{M}}{}_{\underline{\mu}\underline{\nu}}\\
			\delta_\kappa J_a{}_{\bar{\mu}}&=&
			\Delta_a{}^bJ_b{}_{\underline{\mu}}+\partial_a\Delta_{\underline{\mu} }
			+{2}{i}\Delta^{\underline{\nu}}J_a{}^{\underline{M}}
			\gamma_{\underline{M}}{}_{\underline{\mu}\underline{\nu}}	
		\end{array}\right.}}

Under the $\kappa$-symmetry transformation the left sector Lagrangian,
  with $\bar{J}$ and $\bar{\Delta}$ corresponding to $\bar{e}_a{}^m$,
becomes 
\bea
\delta_\kappa {L}|_{\rm{Left}}=\bar{\Delta}_{++} (\bar{J}_-^{\overline{M}})^2
+\bar{\Delta}_{--}(\bar{J}_+^{\overline{M}})^2
+{2}{i}\bar{\Delta}^{\bar{\mu}}\gamma_{\overline{M}}{}_{\bar{\mu}\bar{\nu}} 
\left((1+ k)\bar{J}_+^{\bar{\nu}}\bar{J}_-^{\overline{M}}
+(1- k)\bar{J}_-^{\bar{\nu}}\bar{J}_+^{\overline{M}}\right)
\eea
up to total derivative terms. The Maurer-Cartan equation in \bref{MCLeft} was used.
Let us set the $\kappa$-symmetry parameter for the left sector $\bar{\kappa}$  as
\eq{\bar{\Delta}^{\bar{\mu}}=
	\f12(1+ k)\bar{J}_-^{\overline{M}} \gamma_{\overline{M}}{}^{\bar{\mu}\bar{\nu}}\bar{\kappa}_{+;\bar{\nu}}
	+ 	\f12(1- k)\bar{J}_+^{\overline{M}} \gamma_{\overline{M}}{}^{\bar{\mu}\bar{\nu}}\bar{\kappa}_{-;\bar{\nu}}}
Then it becomes
\bea
\delta_\kappa {L}|_{\rm{Left}}= 
\left(\bar{\Delta}_{++}+{2}{i}(1+k)\bar{\kappa}_{+;\bar{\mu}}\bar{J}_+{}^{\bar{\mu}}\right)(\bar{J}_-^{\overline{M}})^2
+\left(\bar{\Delta}_{--}+{2}{i}(1- k)\bar{\kappa}_{-;\bar{\mu}}\bar{J}_-{}^{\bar{\mu}}\right)(\bar{J}_+^{\overline{M}})^2~~~.
\eea  
Therefore the $\kappa$-symmetry invariance requires the transformation of the zweibein as
\eq{
	0=\bar{\Delta}_{++}+{2}{i}(1+ k)\bar{\kappa}_{+;\bar{\mu}}\bar{J}_+{}^{\bar{\mu}}
	=\bar{\Delta}_{--}+{2}{i}(1-k)\bar{\kappa}_{-;\bar{\mu}}\bar{J}_-{}^{\bar{\mu}}~~~.
}

Analogously the $\kappa$-symmetry transformations for the right sector 
with $\underline{J}$, $\underline{\Delta}$ and $\underline{\kappa}$ are given as
\eq{
	0=\underline{\Delta}_{++}-{2}{i}(1-k)\underline{\kappa}_{+;\underline{\mu}}\underline{J}_+{}^{\underline{\mu}}
	=\underline{\Delta}_{--}-{2}{i}(1+k)\underline{\kappa}_{-;\underline{\mu}}\underline{J}_-{}^{\underline{\mu}}~~~.
}

The Lagrangian $L_1$ is rewritten in a 2D-dimensional covariant way as $L_2$ in \bref{L2}.
Then the manifestly T-duality covariant superstring Lagrangian is given by 
\bea
&&L_2=L_{2:\rm{kin}}+L_{2:\rm{SD}}+L_{2:\rm{WZ}}\nn\\
&&	{\renewcommand{\arraystretch}{1.6}\left\{
	\begin{array}{ccl}
		{L}_{2:\rm{kin}}&=&\displaystyle\frac{1}{{e}}~({e}_+{}^m{J}_m^{{M}})({e}_-{}^n J_n^{{N}}) \hat{\eta}_{{M}{N}}\\
		L_{2:\rm{SD}}&=&
		\displaystyle(\frac{1}{\bar{e}}-\frac{1}{e})({e}_-{}^n J_n^{\overline{M}})^2
		+	(\frac{1}{\underline{e}}-\frac{1}{e})({e}_+{}^n J_n^{\underline{M}})^2 \\
		{L}_{2:\rm{WZ}}&=&
		\displaystyle\frac{k}{2}
		\left(	 J_{[0}{}^{\bar{\mu}} J_{1]}{}_{\bar{\mu}})
		-J_{[0}{}^{\underline{\mu}}J_{1]}{}_{\underline{\mu}}\right)
	\end{array}\right.}\label{Tsust2}
\eea
The kinetic  term has both the 2D-target space covariance and the worldsheet covariance.

Next let us examine how the Lagrangian reduces into the Green-Schwarz superstring Lagrangian.
The left and right Lorentz symmetries can be fixed in such a way that 
two gamma matrices are identified to the D-dimensional gamma matrix as 
 $\gamma^{\overline{M}}{}_{\bar{\mu}\bar{\nu}}=-\gamma^{\underline{M}}{}_{\underline{\mu\nu}}
=\gamma^{\rm M}{}_{\mu\nu}$.
The left and right spinors $\theta^{\bar{\mu}}$ and $\theta^{\underline{\mu}}$ can be chosen to be 
the opposite chirality for the type IIA and the same  chirality for the type IIB theories.
In terms of $x^{\rm{M}},~y_{\rm{M}}$ coordinates in \bref{322}, the currents are written as
\eq{	{\renewcommand{\arraystretch}{1.6}\left\{
		\begin{array}{ccl}
			{\bf J}_m{}^{\rm M}&=&J_m{}^{\overline{M}}+J_m{}^{\underline{M}}=\partial_m x^{\rm M}
			-i(\theta^{\bar{\mu}}\gamma^{\rm M}{}_{\mu\nu}\partial_m\theta^{\bar{\nu}}
			+\theta^{\underline{\mu}}\gamma^{\rm M}{}_{\mu\nu}\partial_m\theta^{\underline{\nu}}		)\\
			{\bf J}_m{}_{\rm M}&=&J_m{}^{\overline{M}}-J_m{}^{\underline{M}}=\partial_m y_{\rm M}
			-i(\theta^{\bar{\mu}}\gamma_{\rm M}{}_{\mu\nu}\partial_m\theta^{\bar{\nu}}
			-\theta^{\underline{\mu}}\gamma_{\rm M}{}_{\mu\nu}\partial_m\theta^{\underline{\nu}}		)\\
			J_m{}^{\bar{\mu}}&=&\partial_m\theta^{\bar{\mu}}\\
			J_m{}^{\underline{\mu}}&=&\partial_m\theta^{\underline{\mu}}\\
			J_{m;\bar{\mu}}&=&\partial_m\varphi_{\bar{\mu}}
			-i\left(2\partial_m (x+y)^{\rm{M}}
			-\frac{2i}{3}\theta^{\bar{\nu}}\gamma^{\rm{M}}{}_{\bar{\nu}\bar{\rho}}\partial_m\theta^{\bar{\rho}}\right)(\theta\gamma_{\rm{M}})_{\bar{\mu}}	\\
			J_{m;\underline{\mu}}&=&\partial_m\varphi_{\underline{\mu}}
			-i\left(2\partial_m(x-y)^{\rm{M}}
			-\frac{2i}{3}\theta^{\underline{\nu}}\gamma^{\rm{M}}{}_{\underline{\nu\rho}}\partial_m\theta^{\underline{\rho}}\right)(\theta\gamma_{\rm{M}})_{\underline{\mu}}		
		\end{array}\right.} \label{JsdJasd}}
Then the  Lagrangian $L_2$ is rewritten as 
\bea
&&L_2=L_{2:\rm{kin}}+L_{2:\rm{SD}}+L_{2:\rm{WZ}}\nn\\
&&	{\renewcommand{\arraystretch}{1.6}\left\{
	\begin{array}{ccl}
		{L}_{2:\rm{kin}}&=&\displaystyle\frac{1}{2{e}}~\left\{
		({e}_+{}^m{\bf J}_m{}^{\rm{M}})({e}_-{}^n {\bf J}_n{}^{\rm{N}}) {\eta}_{\rm{MN}}
		+ ({e}_+{}^m{\bf J}_m{}_{\rm{M}})({e}_-{}^n {\bf J}_n{}_{\rm{N}}) {\eta}^{\rm{MN}}\right\}\\
		L_{2:\rm{SD}}&=&\frac{\lambda}{4e} 
		\left\{ (e_+{}^m+e_-{}^m){\bf J}_m{}^{\rm M}-(e_+{}^m-e_-{}^m){\bf J}_m{}_{\rm M}\right\}\\
&&\times\left\{- (e_+{}^m-e_-{}^m){\bf J}_m{}^{\rm M}+(e_+{}^m+e_-{}^m){\bf J}_m{}_{\rm M}\right\}			\\
		{L}_{2:\rm{WZ}}&=&
		\displaystyle\frac{k}{2}
		\left(	 J_{[0}{}^{\bar{\mu}} J_{1]}{}_{\bar{\mu}})
		-J_{[0}{}^{\underline{\mu}}J_{1]}{}_{\underline{\mu}}\right)
	\end{array}\right.}\label{531}
\eea
with $\lambda=\frac{\lambda_+}{e+\lambda_+}$.
The selfduality constraint $L_{2;{\rm{SD}}}$ in the conformal gauge $g_{\pm}=1$ is given by
\eq{	L_{2:\rm{SD}}=\frac{\lambda}{2} 
( {\bf J}_0{}^{\rm M}-{\bf J}_1{}_{\rm M})
({\bf J}_1{}^{\rm M}-{\bf J}_0{}_{\rm M})}
where the selfduality condition for a superstring is
\eq{{\bf J}_m{}_M=\epsilon_{mn}{\bf J}_n{}^N\eta_{NM}~~~.	} 
The zweibein squared is the worldsheet metric $g^{mn}=e_a{}^me_b{}^n\rho^{ab}$ with 
$-\rho^{00}=1=\rho^{11}$ and $g=\det g_{mn}$. 
Then kinetic term $L_{2:\rm{kin}}$ is given as
\eq{		{L}_{2:\rm{kin}}=-\displaystyle\frac{1}{{2}}\sqrt{g}~\left\{
		g^{mn}{\bf J}_m{}^{\rm{M}}{\bf J}_n{}^{\rm{N}} {\eta}_{\rm{MN}}
		+ g^{mn}{\bf J}_m{}_{\rm{M}}{\bf J}_n{}_{\rm{N}} {\eta}^{\rm{MN}}\right.\}~~~.}
The Wess-Zumino term is given as
\bea
{L}_{2:\rm WZ}&=&
		\displaystyle\frac{k}{2}\epsilon^{mn}
				\left\{(	\partial_m\theta^{\bar{\mu}}\partial_n\varphi_{\bar{\mu}}
				-\partial_m\theta^{\underline{\mu}}\partial_n\varphi_{\underline{\mu}})\right.\nn\\
&&\left.-\partial_m x^{\rm M}~i(\theta^{\bar{\mu}}\partial_n\theta^{\bar{\nu}}- \theta^{\underline{\mu}}\partial_n\theta^{\underline{\nu}})\gamma_{\rm M}{}_{\mu\nu}				
-\partial_m y_{\rm M}~i(\theta^{\bar{\mu}}\partial_n\theta^{\bar{\nu}}+ \theta^{\underline{\mu}}\partial_n\theta^{\underline{\nu}})\gamma_{\rm M}{}_{\mu\nu}	\right\}
\eea

The Green-Schwarz superstring Lagrangian is obtained  from \bref{531} by
the gauge $\lambda=0$ and a section with ${\bf J}_m{}_{\rm M}=0$;
\bea
&&L_2=L_{2:\rm{kin}}+L_{2:\rm{WZ}}\nn\\
&&	{\renewcommand{\arraystretch}{1.6}\left\{
	\begin{array}{ccl}
		{L}_{2:\rm{kin}}&=&-\displaystyle\frac{1}{{2}}\sqrt{g}
		g^{mn}{\bf J}_m{}^{\rm{M}}{\bf J}_n{}^{\rm{N}} {\eta}_{\rm{MN}}\\
		{L}_{2:\rm{WZ}}&=&
		\displaystyle\frac{k}{2}\epsilon^{mn}
	\left\{(	\partial_m\theta^{\bar{\mu}}\partial_n\varphi_{\bar{\mu}}
	-\partial_m\theta^{\underline{\mu}}\partial_n\varphi_{\underline{\mu}})\right.\\
	&&\left.-i\partial_m x^{\rm M}(\theta^{\bar{\mu}}\partial_n\theta^{\bar{\nu}}- \theta^{\underline{\mu}}\partial_n\theta^{\underline{\nu}})\gamma_{\rm M}{}_{\mu\nu}	+		
(\theta^{\bar{\mu}}\gamma^{\rm M}{}_{\mu\nu}\partial_m\theta^{\bar{\nu}})	
( \theta^{\underline{\rho}}\gamma_{\rm M}{}_{\rho\lambda}\partial_n\theta^{\underline{\lambda}})\right\}
	\end{array}\right.}
\eea
The auxiliary fermions $\varphi_\mu$'s appear only in the surface terms 
which are absent in the usual Green-Schwarz superstring action,
so $\varphi_\mu$'s are gauged away.
It is interesting that the bosonic Wess-Zumino term $\epsilon^{mn}{\bf J}_m{}^{\rm M}{\bf J}_n{}_{\rm M}$ does not show up contrasting to the chiral approach \cite{Hatsuda:2015cia}.
 
\section{Conclusions}

In this paper we have presented Lagrangians with manifest T-duality for the type II superstring.
The chiral scalar problem is solved by adding the anti-selfdual currents.
While only the selfdual currents are physical, the unphysical anti-selfdual currents 
are also included for the worldsheet covariance of both the Weyl and Lorentz symmetries.
The selfduality constraints are imposed to suppress the degrees of freedom of 
the anti-selfdual currents whose Lagrange multipliers become the worldsheet zweibeins.  
Then the Lagrangian has double zweibeins and so double worldsheets.
Double zweibeins in the superstring Lagraigian make the type II $\kappa$-symmetry
splitting into two sets of the type I $\kappa$-symmetries leading to simpler computation of $\kappa$-symmetry.
Resulted superstring Lagrangian with manifest T-duality including double zweibeins is
\bref{doublezweibeins}, the one with single zweibein is
\bref{Tsust2} and the one in terms of  the usual $x^{\rm M}, y_{\rm M}$ coordinates with single zweibein is
\bref{531}.
The superstring Lagrangians are simple structure; all terms are given in bilinears of 
symmetry invariant currents manifesting  the global supersymmetry, the T-duality symmetry, coordinate invariance and the $\kappa$-symmetry.
We give the gauge condition and the section condition 
which lead to the Green-Schwarz superstring Lagrangian.

Including the Ramond-Ramond field based on the central extended superalgebra \cite{Hatsuda:2014aza}
and constructing D-brane Lagrangians are future problems.
Superstrings and D-branes with background fields will be also interesting.

\section*{Acknowledgements}

M.H. would like to thank the Simons Center for Geometry and Physics for
hospitality during ``the 2019 Summer Simons workshop in Mathematics and Physics"
where this work has been developed.
W.S. is supported by NSF grant PHY-1620628.

\appendix
\newpage

\section{Indices}
\label{ind}

Indices are summarized.

$$ \vcenter{\halign{ \hfil # : & # \hfil \cr
		$Integer$&Nonabelian space $\cdots I,J,\cdots$ \cr
		{\it Caligraphy}&Superspace $\cdots {\cal M},{\cal N},\cdots$ \cr
		$Middle$ & doubled vector $\cdots M,N,\cdots$ \cr
		{\it greek}&{ doubled spinors}  $\cdots {\mu},{\nu},\cdots$ \cr
		\omit & \cr
		{\it UPPER CASE} & spacetime $\cdots M,N,\cdots$ \cr
		{\it lower case} & worldvolume $\cdots m,n,\cdots$ \cr
		\omit & \cr
		{\it $\bar{B}$arred} & left-handed $\cdots  ~\overline{M},\overline{N},\cdots $ \cr
		{\it \underline{U}nderlined} & right-handed $\cdots  ~\underline{M},\underline{N},\cdots$ \cr
		Roman & {D-dimensional vector}  $\cdots {\rm M},{\rm N},\cdots$ \cr
}} $$
\small
\linespread{1.1}\selectfont
\raggedright

\providecommand{\href}[2]{#2}\begingroup\raggedright\endgroup

\end{document}